\documentclass[11pt, letterpaper]{article}

\usepackage{ifthen}
\newcommand{\Bset}[1]{\setboolean{#1}{true}}
\newcommand{\Bunset}[1]{\setboolean{#1}{false}}

\usepackage{amssymb}
\usepackage{amsmath}   %
\usepackage{amsthm}   %
\usepackage{latexsym}
\usepackage{amsfonts}
\usepackage{psfrag}    %
\usepackage{url}    %

\usepackage[dvips]{graphicx}
\usepackage[usenames,dvipsnames]{color}

\newcommand{\Bif}[3]{\ifthenelse{\boolean{#1}}{#2}{#3}}

{\theoremstyle{remark}}

\newtheoremstyle{mydefinition}   %
{\topsep}{\topsep}   %
{\slshape}   %
{}   %
{\bfseries}   %
{.}   %
{ }   %
{}   %
{\theoremstyle{mydefinition}}
\newtheoremstyle{myexample}   %
{\topsep}{\topsep}   %
{\itshape}   %
{}   %
{\slshape}   %
{:}   %
{ }   %
{\ul{\thmname{#1}}}   %
{\theoremstyle{myexample}}

\newtheoremstyle{anystatement}{\topsep}{\topsep}{\itshape}{}{\bfseries}{.}{ }{\anystatementname}
{\theoremstyle{anystatement}}

\newcommand{\anystatementname}{}

\newcounter{tmp_id_cnt}
\newcommand{\nospell}[1]{#1}   %

\newcommand{\mydef}[2]{\def#1{#2}}

\newcommand{\newident}[3][*]{\ifthenelse{\equal{*}{#1}}      %
{\newcommand{#2}[1][*]                                     %
{\ifthenelse{\equal{*}{##1}}                              %
{\nospell{\mbox{\Ensuremath{{\mathit{#3}}}}}}            %
{\ifthenelse{\equal{b}{##1}}                             %
{\nospell{\mbox{\Ensuremath{{\mathbf{#3}}}}}}          %
{#3}}}}                                                %
{\mydef{#2}{\nospell{\mbox{\Ensuremath{{\mathit{#3}}}}}}}}

\newcommand{\newidentarg}[2]{                                                   %
\newcommand{#1}[1][]                                                          %
{\ifthenelse{\boolean{in_math_mode}}                                          %
{#2}                                                                        %
{\nospell{\mbox{\Ensuremath{\mathit{#2}}}}}}}                               %

\newcommand{\newmat}[3][*]{\ifthenelse{\equal{*}{#1}}       %
{\newcommand{#2}[1][*]                                           %
{\ifthenelse{\equal{b}{##1}}                                    %
{\nospell{\mbox{\Ensuremath{\mathbf{#3}}}}}                    %
{\ifthenelse{                                                  %
\( \equal{*}{##1} \and \not \boolean{in_math_mode} \)         %
\or \( \not \equal{*}{##1} \and \boolean{in_math_mode} \)}    %
{\nospell{\mbox{\Ensuremath{#3}}}}                            %
{#3}}}}                                                       %
{\mydef{#2}{#3}}}                                                %

\newcommand{\newmatarg}[2]{                               %
\newcommand{#1}[1][]                                                   %
{\ifthenelse{\boolean{in_math_mode}}                                  %
{#2}                                                                %
{\nospell{\mbox{\Ensuremath{#2}}}}}}                                %

\newcommand{\newmatop}[2]{\mydef{#1}{\operatorname{#2}}}

\newcommand{\MyMakeTheoMacros}[3]{
\newcommand{#2}[2][]{\ifthenelse{\equal{}{##1}}
{\begin{#1} ##2 \end{#1}}
{\begin{#1}\label{##1} ##2\end{#1}}}
\newcommand{#3}[3][]{\ifthenelse{\equal{}{##1}}
{\begin{#1}{\e{##2}} ##3 \end{#1}}
{\begin{#1}{\e{##2}}\label{##1} ##3\end{#1}}}
}

\newcommand{\MyMakeDupTheoMacros}[8]{
\MyMakeTheoMacros{#1}{#2}{#3}
\newcommand{#4}[3]{
\newcommand{##2}{##3}
\begin{#1}\label{##1} ##2\end{#1}}
\newcommand{#5}[4]{
\newcommand{##2}{##4}
\begin{#1}{\e{##3}}\label{##1} ##2\end{#1}}
\newcommand{#8}[2]{\def\my_tmp_id{my_tmp_id_\arabic{tmp_id_cnt}}
\newtheorem*{\my_tmp_id}{#7~\ref{##1}}
\begin{\my_tmp_id} ##2 \end{\my_tmp_id}\stepcounter{tmp_id_cnt}}
\newcommand{#6}[6]{
#2[##1]{##2}

##3
\prf[#7~\ref{##1}]{##6} \newcommand{##5}{}

}
}

\newcommand{\MyMakeRefMacros}[3]{\newcommand{#1}[2][]
{\ifthenelse{\equal{}{##1}}{#2~\ref{##2}}{#3~\ref{##1} and~\ref{##2}}}}

\newcommand{\MyMakeEqRefMacros}[3]{\newcommand{#1}[2][]
{\ifthenelse{\equal{}{##1}}{#2~\eqref{##2}}{#3~\eqref{##1} and~\eqref{##2}}}}

\newcommand{\abstr}[1]{
\begin{abstract}
#1
\end{abstract}}

\newcommand{\bibentry}[8]{

\bibitem[\nospell{#8}]{#1} {\textup #3}. 
\ifthenelse{\equal{}{#6}}
{\newblock \textrm{#4.} \newblock {\em #5}, #7.}
{\newblock \textrm{#4.} \newblock {\em #5, #6}, #7.}
}
\newcommand{\bibentryQ}[8]{

\bibitem[\nospell{#8}]{#1} {\textup #3}. 
\ifthenelse{\equal{}{#6}}
{\newblock \textrm{#4} \newblock {\em #5}, #7.}
{\newblock \textrm{#4} \newblock {\em #5, #6}, #7.}
}

\newcommand{\inputbib}{

\bibentryPerCom[2006]{H. Buhrman}{B}

\bibentry{B64_On}{Bell}{J. Bell}{On the Einstein-Podolsky-Rosen Paradox}{Physics 1(3)}{pages 195-200}{1964}{B64}

\bibentry{BCW98}{Buhrman, Cleve and Wigderson}{H. Buhrman, R. Cleve and A. Wigderson}{Quantum vs. Classical Communication and Computation}{Proceedings of the 30th Symposium on Theory of Computing}{pages 63-68}{1998}{BCW98}

\bibentry{BCWW01}{Buhrman, Cleve, Watrous and de Wolf}{H. Buhrman, R. Cleve, J. Watrous and R. de Wolf}{Quantum Fingerprinting}{Physical Review Letters 87(16)}{article 167902}{2001}{BCWW01}

\bibentry{BHMR03}{Buhrman, H{\o}yer, Massar and R{\"o}hrig}{H. Buhrman, P. H{\o}yer, S. Massar and H. R{\"o}hrig}{Combinatorics and Quantum Nonlocality}{Physical Review Letters 91}{article 047903}{2003}{BHMR03}

\bibentry{BHMR06}{Buhrman, H{\o}yer, Massar and R{\"o}hrig}{H. Buhrman, P. H{\o}yer, S. Massar and H. R{\"o}hrig}{Multipartite Nonlocal Quantum Correlations Resistant to Imperfections}{Physical Review Letters A 73}{article 012321}{2006}{BHMR06}

\bibentry{BJK04}{Bar-Yossef, Jayram and Kerenidis}{Z. Bar-Yossef, T. S. Jayram and I. Kerenidis}{Exponential Separation of Quantum and Classical One-Way Communication Complexity}{Proceedings of 36th Symposium on Theory of Computing}{pages 128-137}{2004}{BJK04}

\bibentryPerCom[2005]{R. Cleve}{C}

\bibentry{CHTW04}{Cleve, H{\o}yer, Toner and Watrous}{R. Cleve, P. H{\o}yer, B. Toner and J. Watrous}{Consequences and Limits of Nonlocal Strategies}{Proceedings of the 19th IEEE Conference on Computational Complexity}{pages 236-249}{2004}{CHTW04}

\bibentryQ{EPR35}{Einstein, Podolsky and Rosen}{A. Einstein, B. Podolsky and N. Rosen}{Can Quantum-Mechanical Description of Physical Reality be Considered Complete?}{Physical Review 41}{pages 777-780}{1935}{EPR35}

\bibentry{G08_C_I_C_R}{Gavinsky}{D. Gavinsky}{Classical Interaction Cannot Replace a Quantum Message}{Proceedings of the 40th Symposium on Theory of Computing}{pages 95-102}{2008}{G08}

\bibentry{GKKRW07}{Gavinsky, Kempe, Kerenidis, Raz and de Wolf}{D. Gavinsky, J. Kempe, I. Kerenidis, R. Raz and R. de Wolf}{Exponential Separations for One-Way Quantum Communication Complexity, with Applications to Cryptography}{Proceedings of the 39th Symposium on Theory of Computing}{pages 516-525}{2007}{GKKRW07}

\bibentry{GKRW06}{Gavinsky, Kempe, Regev and de Wolf}{D. Gavinsky, J. Kempe, O. Regev and R. de Wolf}{Bounded-error Quantum State Identification and Exponential Separations in Communication Complexity}{Proceedings of the 38th Symposium on Theory of Computing}{pages 594-603}{2006}{GKRW06}

\bibentry{HJAR07}{Harsha, Jain, McAllester and Radhakrishnan}{P. Harsha, R. Jain, D. McAllester and J. Radhakrishnan}{The communication complexity of correlation}{Proceedings of the 22nd IEEE Conference on Computational Complexity}{pages 10-23}{2007}{HJAR07}

\bibentry{R99}{Raz}{R. Raz}{Exponential Separation of Quantum and Classical Communication Complexity}{Proceedings of the 31st Symposium on Theory of Computing}{pages 358-367}{1999}{R99}

\bibentry{Y79}{Yao}{A. C-C. Yao}{Some Complexity Questions Related to Distributed Computing}{Proceedings of the 11th Symposium on Theory of Computing}{pages 209-213}{1979}{Y79}

}

\newcommand{\bib}[1][]{

\begin{thebibliography}{ABCDEFG00}
\frenchspacing
\inputbib
#1
\end{thebibliography}
}

\newcommand{\citePerCom}[2][]{\cite{#2}}
\newcommand{\bibentryPerCom}[3][]{ \ifthenelse{\equal{}{#1}}
{
\bibitem[\nospell{#3}]{#2} {\textup #2} - \textit{Personal communication}. }
{
\bibitem[\nospell{#3}]{#2} {\textup #2} - \textit{Personal communication}, #1. 
}
}

\MyMakeTheoMacros{fact}{\fct}{\nfct}

\MyMakeRefMacros{\fctref}{Fact}{Facts}

\MyMakeTheoMacros{observation}{\obs}{\nobs}

\MyMakeRefMacros{\obsref}{Observation}{Observations}

\MyMakeDupTheoMacros{lemma}
{\lem}{\nlem}{\lemdup}{\nlemdup}{\lemapp}{Lemma}{\lemrep}

\MyMakeRefMacros{\lemref}{Lemma}{Lemmas}

\newcommand{\fakelemref}[1]{Lemma~{#1}}

\MyMakeDupTheoMacros{corollary}
{\crl}{\ncrl}{\crldup}{\ncrldup}{\crlapp}{Corollary}{\crlrep}

\MyMakeRefMacros{\crlref}{Corollary}{Corollaries}

\MyMakeTheoMacros{proposition}{\prp}{\nprp}

\newtheorem*{prp*}{\e{Proposition}}
\newcommand{\prpunnambered}[1]{\begin{prp*} #1 \end{prp*}}
\mydef{\uprp}{\prpunnambered}

\MyMakeRefMacros{\prpref}{Proposition}{Propositions}

\MyMakeDupTheoMacros{claim}
{\clm}{\nclm}{\clmdup}{\nclmdup}{\clmapp}{Claim}{\clmrep}

\MyMakeRefMacros{\clmref}{Claim}{Claims}

\MyMakeDupTheoMacros{theorem}
{\theo}{\ntheo}{\theodup}{\ntheodup}{\theoapp}{Theorem}{\theorep}

\MyMakeRefMacros{\theoref}{Theorem}{Theorems}

\MyMakeTheoMacros{definition}{\defi}{\ndefi}

\MyMakeRefMacros{\defiref}{Definition}{Definitions}

\MyMakeTheoMacros{problem}{\prob}{\nprob}

\MyMakeRefMacros{\probref}{Problem}{Problems}

\MyMakeTheoMacros{protocol}{\prot}{\nprot}

\MyMakeRefMacros{\protref}{Protocol}{Protocols}

\providecommand{\qedsymbol}{\square}

\newcommand{\prf}[2][]{\ifthenelse{\equal{}{#1}}                           %
{\begin{proof}\renewcommand{\qedsymbol}{$\blacksquare$}                  %
#2 \end{proof}}                                                       %
{\begin{proof}[Proof of #1]                                               %
\renewcommand{\qedsymbol}{$\blacksquare_{\mbox{\it{\scriptsize{#1}}}}$}  %
#2 \end{proof}}
}

\newcommand{\sect}[2][]{\ifthenelse{\equal{}{#1}}
{\section{#2}}
{\section{#2}\label{#1}}}

\newcommand{\ssect}[2][]{\ifthenelse{\equal{}{#1}}
{\subsection{#2}}
{\subsection{#2}\label{#1}}}

\MyMakeRefMacros{\chref}{Chapter}{Chapters}

\MyMakeRefMacros{\sref}{Section}{Sections}

\MyMakeRefMacros{\ssref}{Subsection}{Subsections}

\MyMakeRefMacros{\sssref}{Subsection}{Subsections}

\definecolor{DarkGreen}{rgb}{0,0.45,0.08}

\newboolean{in_math_mode}
\setboolean{in_math_mode}{false}

\newcommand{\IfMathMode}{\Bif{in_math_mode}}

\newcommand{\MathModeOn}{\Bset{in_math_mode}}

\newcommand{\MathModeOff}{\Bunset{in_math_mode}}

\newcommand{\Ensuremath}[1]{\IfMathMode   %
{\ensuremath{#1}}
{\MathModeOn\ensuremath{#1}\MathModeOff}}

\newcommand{\fbr}[1]{\IfMathMode   %
{$#1$}                              %
{\MathModeOn$#1$\MathModeOff}}

\newcommand{\fnbr}[1]{\mbox{\fbr{#1}}}   %

\newcommand{\fla}[2][*]{\ifthenelse{\equal{}{#1}}{\fbr{#2}}{\fnbr{#2}}}

\newcommand{\mat}[2][]{\ifthenelse{\equal{}{#1}}   %
{\begin{displaymath} \MathModeOn
#2
\MathModeOff \end{displaymath}     %
}{
\begin{equation} \MathModeOn \label{#1}
#2
\MathModeOff \end{equation}
}
}

\newcommand{\matal}[2][]{\mat[#1]{\begin{split} #2 \end{split}}}

\newcommand{\f}{\fla}

\newcommand{\m}{\mat}

\newcommand{\mal}{\matal}

\newcommand{\mac}{\substack}    %

\MyMakeEqRefMacros{\equref}{Equation}{Equations}

\MyMakeEqRefMacros{\expref}{Expression}{Expressions}

\MyMakeEqRefMacros{\inequref}{Inequality}{Inequalities}

\newcommand{\bracref}[1]{(\ref{#1})}

\newcommand{\bref}{\bracref}

\MyMakeRefMacros{\figref}{Figure}{Figures}

\providecommand{\middle}{\big}

\newmatop{\mymod}{mod}   %
\newmatop{\poly}{poly}
\newmatop{\sign}{sign}
\newmatop{\tr}{tr}
\newmatop{\supp}{supp}    %
\newmatop{\argmax}{argmax}
\newmatop{\argmin}{argmin}
\newmatop{\posmin}{posmin}
\newmatop{\negmin}{negmin}
\newmatop{\posmax}{posmax}
\newmatop{\negmax}{negmax}
\newmatop{\argmaxmin}{argmax(min)}

\newcommand{\PR}[2][]{\mathop{\mathbf{Pr}}_{#1}\left[{#2}\right]}
\newcommand{\PRr}[3][]{\mathop{\mathbf{Pr}}_{#1}\left[{#2}\middle|\vphantom{|_1^1}{#3}\right]}

\newcommand{\U}[1][]{\ifthenelse{\equal{}{#1}}
{{\cal U}}
{{\cal U}_{#1}}}
\newcommand{\Uu}[2]{\ifthenelse{\equal{}{#1}}
{{\cal U}^{#2}}
{{\cal U}_{#1}^{(#2)}}}

\newcommand{\GF}[1]{{\cal GF}_{#1}}

\newcommand{\pss}[1][]{\nospell{\ifthenelse{\equal{}{#1}}
{\mbox{'s}}
{\fla{#1}\mbox{'s}}}}

\newcommand{\pl}[1][]{\nospell{\ifthenelse{\equal{}{#1}}
{\mbox{-s}}
{\fla{#1}\mbox{-s}}}}

\newcommand{\ord}[1][]{\nospell{\ifthenelse{\equal{}{#1}}
{\mbox{'th}}
{\ifthenelse{\equal{1}{#1}}{$1$\mbox{'st}}{\ifthenelse{\equal{2}{#1}}{$2$\mbox{'nd}}{\ifthenelse{\equal{3}{#1}}{$3$\mbox{'rd}}{\fla{#1}\mbox{'th}}}}}}}

\newmat[]{\llp}{\left(}
\newmat[]{\rrp}{\right)}

\newmat[]{\lla}{\left\langle}
\newmat[]{\rra}{\right\rangle}

\newmat[]{\llc}{\left\lceil}
\newmat[]{\rrc}{\right\rceil}

\newmat[]{\llf}{\left\lfloor}
\newmat[]{\rrf}{\right\rfloor}

\newmat[]{\tm}{\cdot}
\newmat[]{\sbseq}{\subseteq}
\newmat[]{\sbs}{\subset}
\newmat[]{\smin}{\setminus}
\newmat[]{\eps}{\varepsilon}
\newmat[]{\deq}{\stackrel{\textrm{def}}{=}}

\newmat{\NN}{\mathbb{N}}
\newmatarg{\ZZ}{\ifthenelse{\equal{}{#1}}{\mathbb{Z}}{\mathbb{Z}_{#1}}}

\newmat[]{\ds}{\dots}
\newmat[]{\dc}{,\ds,}

\newmat[]{\sdnle}{\npreceq}

\newcommand{\enum}[1]{\begin{enumerate} #1 \end{enumerate}}

\newcommand{\itemi}[2][]{\ifthenelse{\equal{}{#1}}
{\begin{itemize} #2 \end{itemize}}
{\begin{itemize}[#1] #2 \end{itemize}}}

\newcommand{\wrt}{w.r.t.\ }	 %

\newcommand{\ie}{i.e., }	 %

\newcommand{\fr}[3][*]{
\ifthenelse{\equal{*}{#1}}               %
{\frac{#2}{#3}}{}
\ifthenelse{\equal{}{#1}}                %
{\left.#2\middle/#3\right.}{}
\ifthenelse{\equal{b_}{#1}}              %
{\left.\left(#2\right)\middle/#3\right.}{}
\ifthenelse{\equal{_b}{#1}}              %
{\left.#2\middle/\left(#3\right)\right.}{}
\ifthenelse{\equal{bb}{#1}}              %
{\left.\left(#2\right)\middle/\left(#3\right)\right.}{}
}

\newcommand{\set}[2][]{\ifthenelse{\equal{}{#1}}               %
{\Ensuremath{\left\{#2\right\}}}                             %
{\Ensuremath{\left\{#2\middle|\vphantom{|_1^1}#1\right\}}}}  %

\newcommand{\Min}[2][]{\ifthenelse{\equal{}{#1}}   %
{\Ensuremath{\min\left\{#2\right\}}}             %
{\Ensuremath{\min_{#1}\left\{#2\right\}}}}       %

\newcommand{\Max}[2][]{\ifthenelse{\equal{}{#1}}   %
{\Ensuremath{\max\left\{#2\right\}}}             %
{\Ensuremath{\max_{#1}\left\{#2\right\}}}}       %

\newcommand{\newfunction}[2]{                          %
\newcommand{#1}[2][*]{\ifthenelse{\equal{*}{##1}}    %
{\Ensuremath{#2\left(##2\right)}}                   %
{#2(##2)}}                                          %
}                                                     %

\newfunction{\asO}{O}
\newfunction{\astO}{\tilde O}
\newfunction{\aso}{o}
\newfunction{\asOm}{\Omega}
\newfunction{\astOm}{\tilde \Omega}
\newfunction{\asom}{\omega}
\newfunction{\asT}{\Theta}

\newcommand{\ket}[1]{\Ensuremath{\left|#1\rra}}
\newcommand{\bra}[1]{\Ensuremath{\lla #1\right|}}
\newcommand{\kbra}[2][]{\ifthenelse{\equal{}{#1}}
{\Ensuremath{\left|#2\rra\hspace{-3.5pt}\lla #2\right|}}
{\Ensuremath{\left|#1\rra\hspace{-3.5pt}\lla #2\right|}}
}

\newcommand{\sz}[2][]{\ifthenelse{\equal{}{#1}}
{\Ensuremath{\left|#2\right|}}
{\Ensuremath{\left|#2\right|_{#1}}}}

\newcommand{\lra}[2][*]{\ifthenelse{\equal{}{#1}}
{\langle #2 \rangle}
{\lla #2 \rra}}
\newcommand{\ceil}[2][*]{\ifthenelse{\equal{}{#1}}
{\lceil #2 \rceil}
{\llc #2 \rrc}}
\newcommand{\floor}[2][*]{\ifthenelse{\equal{}{#1}}
{\lfloor #2 \rfloor}
{\llf #2 \rrf}}

\newcommand{\fn}{\footnote}

\newcommand{\nin}{\not\in}   %

\newcommand{\e}{\emph}

\newcommand{\ul}[1]{\underline{#1}}  %

\newcommand{\txt}[1]{\textrm{#1}}   %

\newident{\R}{\mathcal R}
\newident{\RI}{\mathcal R^{1}}
\newident{\RII}{\mathcal R^{\parallel}}
\newident{\RIIp}{\mathcal R^{\parallel,pub}}
\newident{\RIIe}{\mathcal R^{\parallel,ent}}
\newident{\Q}{\mathcal Q}
\newident{\QI}{\mathcal Q^{1}}
\newident{\QII}{\mathcal Q^{\parallel}}
\newident{\QIIp}{\mathcal Q^{\parallel,pub}}
\newident{\QIIe}{\mathcal Q^{\parallel,ent}}

\usepackage{hyphenat}   %

\newcommand{\whereproofs}{Appendix}

\date{}
\newmatop{\cell}{cell}
\newmatop{\infl}{infl}

\newident{\Pnn}{P}
\newidentarg{\Pii}{P_{1\times 1}}

\newmat{\Unn}{\U[n]}
\newmat{\Unp}{\U_n'}
\newmat{\Uii}{\Uu{n}{1\times 1}}

\newmatarg{\UA}{\ifthenelse{\equal{}{#1}}
{\U[A]}
{\Uu{A}{(#1)}}}
\newmatarg{\UAa}{\ifthenelse{\equal{}{#1}}
{\Uu{A}{\A}}
{\Uu{A}{(\A;#1)}}}
\newmatarg{\UAb}{\ifthenelse{\equal{}{#1}}
{\Uu{A}{\B}}
{\Uu{A}{(\B;#1)}}}
\newmatarg{\UB}{\ifthenelse{\equal{}{#1}}
{\U[B]}
{\Uu{B}{(#1)}}}

\newmatarg{\pt}{\ifthenelse{\equal{}{#1}}
{p_t^{(t)}}
{p_{#1}^{(t)}}}

\title{Classical Interaction Cannot Replace Quantum Nonlocality}

\author{
{\bf Dmitry Gavinsky}\\
{\small NEC Laboratories America, Inc.}\\
{\small 4 Independence Way, Suite 200}\\
{\small Princeton, NJ 08540, U.S.A.}
}

\begin{document}

\maketitle

\thispagestyle{empty}

\abstr{We present a two-player communication task that can be solved by a protocol of polylogarithmic cost in the \e{simultaneous} message passing model with classical communication and shared entanglement, but requires exponentially more communication in the classical \e{interactive} model.

Our second result is a two-player nonlocality game with input length $n$ and output of polylogarithmic length, that can be won with probability $1-\aso1$ by players sharing polylogarithmic amount of entanglement.
On the other hand, the game is lost with probability \asOm1 by players without entanglement, even if they are allowed to exchange up to $k$ bits in interactive communication for certain $k\in\astOm{n^{1/4}}$.

These two results give almost the strongest possible (and the strongest known) indication of nonlocal properties of two-party entanglement.}

\setcounter{page}{1}

\sect{Quantum vs.\ classical communication and games}
The ultimate goal of quantum computing is to understand what advantages are offered by using the laws of quantum mechanics to build computational devices.
We want to find tasks where quantum computers are considerably more efficient than classical ones.

In this paper we study quantum computation from the perspective of Communication Complexity, first defined by Yao~\cite{Y79}.
Two parties, Alice and Bob, try to solve a computational problem that depends on $x$ and $y$.
Initially Alice knows only $x$ and Bob knows only $y$; in order to solve the problem they communicate, obeying to restrictions of a specific \e{communication model}.

A \e{communication protocol} describes behavior of the participants in response to each possible input.
The \e{cost} of a protocol is the maximum total amount of (qu)bits communicated by the parties, according to the protocol.

We say that a communication task $P$ is solvable \e{with bounded error} in a given communication model by a protocol of cost \asO{k} if for any constant $\eps>0$, there exists a corresponding protocol solving $P$ with success probability at least $1-\eps$ and communicating \asO{k} (qu)bits.
If the protocols, in addition, either refuse to answer or succeed, then we say that the solution is \e{\f0-error}.

In order to compare the power of two communication models, one has to either prove existence of a task that can be solved more efficiently in one model than in the other, or to argue that no such task exists.

We will, in the first place, be concerned about the following models.\itemi{
\item \e{Interactive (two-way) communication} is a model where the players can interactively exchange messages till Bob decides to give an answer, based on the communication transcript and his part of input.
\item \e{One-way communication} is a model where Alice sends a single message to Bob who has to give an answer, based on the content of the message and his part of input.
\item \e{Simultaneous Message Passing (SMP)} is a model involving third participant, a \e{referee}.
Here both Alice and Bob send one message each to the referee, who has to give an answer, based on the content of the received messages.
}

All three models can be either \e{classical} or \e{quantum}, according to the nature of communication allowed between the players.
The classical versions of the models are denoted by \R, \RI, \RII, and the quantum versions are denoted by \Q, \QI, \QII, respectively.

It is clear that interactive communication is at least as powerful as one-way communication, and it is well-known that the former can, in fact, be much more efficient than the latter, both in quantum and in classical versions.
All the same is true regarding one-way communication vs.\ SMP. 

The model of SMP (both \QII\ and \RII) can be made stronger by allowing some \e{shared resource} between Alice and Bob, which can be either \e{shared randomness} or \e{shared entanglement}.
The former case can be viewed as allowing Alice and Bob to follow \e{mixed} joint strategy, and the latter case means that they can perform quantumly-nonlocal (\ie entangled) operation, as described by the laws of quantum mechanics.
Of course, shared entanglement is always at least as helpful as shared randomness, and it has been shown (\cite{GKRW06}) that the latter can, in fact, be much more efficient than the former (both with \QII\ and with \RII).
When enhanced by shared randomness, the models are denoted as \QIIp\ and \RIIp, and the case of shared entanglement is addressed by \QIIe\ and \RIIe.

In this paper our primary concern is with separating communication models.
Both for previously known results (mentioned later) and for our contribution it is the case that the first demonstrated super-polynomial separation had, in fact, been exponential.

Communication tasks can be either \e{functional}, meaning that there is exactly one correct answer corresponding to every possible input, or \e{relational}, when multiple correct answers are allowed.
Functional tasks over domains forming product sets \wrt each player's inputs are called \e{total}.
The three types of communication tasks form a hierarchy, if viewed as tools to separate communication model.
In particular, there are known pairs of communication models that can be separated through a relational problem but are equally strong over functions, either total or partial, and there are pairs of communication models that can be separated through a partial functional problem but are conjectured to be equally strong over total functions.

For \f0-error one-way and interactive protocols, separations have been demonstrated by Buhrman, Cleve and Wigderson~\cite{BCW98}.
In the bounded-error setting the first separation has been given by Raz~\cite{R99}, showing a problem solvable in \Q\ exponentially more efficiently than in~\R.
Later, Buhrman, Cleve, Watrous and de Wolf~\cite{BCWW01} demonstrated an exponential separation for the SMP model.
All these separations have been given for functional problems.

For one-way protocols with bounded error, the first separation has been shown by Bar-Yossef, Jayram and Kerenidis~\cite{BJK04} for a relational problem.
Later, Gavinsky, Kempe, Kerenidis, Raz and de Wolf~\cite{GKKRW07} gave a similar separation for a partial functional problem.

These results show that quantum communication models can be very efficient, when compared to their classical counterparts.
Recently it has been shown in~\cite{G08_C_I_C_R} that there exists a problem that can be solved by a quantum one-way protocol more efficiently than by any classical two-way protocol.
That result simultaneously subsumes the separation in~\cite{BJK04} and partially that in~\cite{R99} (in \cite{G08_C_I_C_R} we were only able to show separation through a relational problem).

Closely related to SMP models \RIIp\ and \RIIe\ is the notion of \e{nonlocality games}.
In this setting two players, Alice and Bob, receive two parts of input and must produce their output without communicating to each other.
The difference from the SMP setting is that there is no referee (who would help the players to achieve their goal), instead there is a \e{verifier} who checks whether the answers from the players are good \wrt the given input.
Every nonlocality game defines which input pairs are allowed and which pairs of answers are good for each input pair.

There is a number of known nonlocality games (cf.~\cite{CHTW04}) that can be played with higher success probability by entangled (\e{nonlocal}) players than by unentangled (\e{local}) players who share randomness, hence the name.
Historically (\cite{EPR35}, \cite{B64_On}), the setting of nonlocality games has been viewed as a way to demonstrate nonlocal behavior of quantum entanglement.
As a tool for such purpose, it is more powerful (\ie it allows more flexibility and thus is less ``convincing'') than separating \RIIp\ from \RIIe\ through any type of communication task.

\ssect{Our results}
In this work we give even more surprising demonstration of ``usefulness'' of quantum mechanics in solving communication problems.
We give a communication problem that is\itemi{
\item easy in the weakest known communication model with entanglement, \RIIe;
\item hard in (one of) the strongest classical 2-party models, \R.
}
Moreover, our communication problem is also easy in \QI, and therefore this result subsumes~\cite{G08_C_I_C_R}.

\theo[theo_main]{For infinitely many $n\in\NN$, there exists an (explicit) relation with input length $n$ that can be efficiently solved in \RIIe, namely there is an SMP protocol with error \aso1 where the two players share \asO{\log^2n\log\log n} EPR pairs and send classical messages of length \asO{\log^2n\log\log n}.
The same relation has communication complexity \asOm{\fr{n^{1/4}}{\log^2n}} in the interactive classical model (\R).}

The relation we use is a modification of a construction independently suggested by R.\ Cleve (\citePerCom[2005]{R. Cleve}) and S.\ Massar (\citePerCom[2006]{H. Buhrman}), as a possible candidate for such separation.
In~\cite{G08_C_I_C_R} a more ``demanding'' version of this relation was used.
In particular, the communication task from~\cite{G08_C_I_C_R} is unlikely to admit an efficient solution in \RIIe.

Our second result is a new, stronger type of a nonlocality game.
Besides admitting winning probability $1-\aso1$ by two players who share entanglement, but not by any local mixed strategy, the game \e{remains hard even if we allow considerable amount of classical communication between the classical players}.

\theo[theo_nonloc]{For infinitely many $n\in\NN$, there exists an (explicit) 2-party nonlocality game where both players receive input of length $n$ and are required to produce output of length \asO{\log^2n\log\log n}.
The game can be won with probability $1-\aso1$ by players who share \asO{\log^2n\log\log n} bits of entanglement (EPR pairs), but any local mixed strategy would result in loss with probability \asOm1; moreover, the loss would remain \asOm1 even if the players are allowed to interactively exchange \aso{\fr{n^{1/4}}{\log^2n}} classical bits before producing their output.}

The statement is straightforward modulo \theoref{theo_main}:\fn
{Note that no statement in the flavor of \theoref{theo_nonloc} can be derived from~\cite{G08_C_I_C_R}, as the communication task considered there is not known to have an efficient solution in \RIIe.}
\prf{Let $P$ be the communication problem used in \theoref{theo_main}, and let $S$ be an \RIIe-protocol for $P$, as promised.
W.l.g., we assume that the referee's action in $S$ is deterministic (at most \asO{\log n} bits of randomness are required, which can be chosen by Alice and attached to her message).

Let a game $G$ be as follows:\itemi{
\item input to $G$ is the same as input to $P$;
\item for any $(x,y)$, the set of valid responses in $G$ coincides with the set of pairs of messages that would result, according to $S$, in a correct answer by the referee to $(x,y)$.}
Defined this way, $G$ can be won with probability $1-\aso1$ by two entangled players obeying $S$.
On the other hand, $G$ is lost with probability \asOm1 by any classical players exchanging \aso{\fr{n^{1/4}}{\log^2n}} bits, as otherwise they would be able to solve $P$ (in collaboration with a referee obeying $S$), in doing so contradicting the statement of \theoref{theo_main}.}

To the best of author's knowledge, this is the first example of a nonlocality game where an entangled strategy outperforms any classical one, even if the latter is ``boosted'' by allowing up to $k$ bits of interaction, where $k$ is exponential in the total length of players' answers.\fn
{Examples of \asOm{n}-party nonlocality games efficiently played with entanglement, where classical solution would require $n^{\asOm1}$ bits of auxiliary communication were given by Buhrman, H{\o}yer, Massar and R{\"o}hrig in~\cite{BHMR03} and~\cite{BHMR06} -- of course, the total output length is also \asOm{n} in that case.}
The fact that our game is a simple 2-party one makes it an appealing candidate for an actual physical experiment.

\sect{Previous and new techniques}
The main result of this paper can be viewed as strengthening of~\cite{G08_C_I_C_R}.
The communication task analyzed in this paper is arguably more basic than that from~\cite{G08_C_I_C_R}, and its analysis requires more refined tools.
On the high level, the difference can be viewed as follows.

Ignoring some technical details,\fn
{In the actual construction we allow certain deviations from this description, as required for our proof technique to go through.}
in the both cases the task is:\itemi{
\item there is an $n\times n$ table, where each cell contains $2$ elements;
\item every element from \set{1\dc2n^2} appears in the table once;
\item Alice knows the row and Bob knows the column of each element;
\item for the answer the players must choose a cell and give certain witness of knowledge about the both element in it.}
The intuitive motivation to ask for a witness is that \e{ability to provide it would implicitly require from classical players to know the element of the chosen cell}, while entangled players are able to get a witness using much less communication than it would be required in order to learn the content of a cell.

The two versions of the communication task differ in \e{how may the players choose a cell for their answer}.
In this paper we consider a version where the players can pick \e{any cell}.
In~\cite{G08_C_I_C_R} the players had to pick a \e{cell from the first row}.

This (apparently minor) difference surprisingly makes the earlier version considerably harder to solve; in particular, it is unlikely to admit an efficient solution in the model of simultaneous messages with entanglement (but still has an efficient solution in the model of quantum one-way communication, as demonstrated in~\cite{G08_C_I_C_R}).
The same difference makes the earlier version less ``resistant'' to lower bound techniques.

The lower bound proof, in the both cases, consists of several ``quasi-reductions'', the first of which can be (again, ignoring some technicalities) stated as 
\uprp{If there is a two-way classical solution of cost $k$ to the communication task under consideration, then the $1\times1$-version of the task can be solved with probability \asOm{1/n} by a protocol of similar cost.}
In this statement by a $1\times1$-version we mean a modification of the task where the players are more restricted -- namely, they no longer have freedom to choose a cell, but must give their answer \wrt \e{the first cell of the first row}.

We can see that the proposition is almost trivial if the communication task under consideration is the earlier version:
The first row has $n$ elements, so with probability $1/n$ ``a cell from the first row'' will turn out to be the first one (of course, an adversarial scenario would be that a protocol for the original problem never outputs an answer based on the first cell, but that can be helped by randomly permuting the columns in the beginning).

When we give the players more freedom and let them choose any cell from the table, we make the problem affordable for the model of simultaneous messages with entanglement, but the aforementioned proof idea is no longer valid:\ there are $n^2$ cells and a similar reduction would only guarantee a solution for the $1\times1$-version with probability $1/n^2$ (which is easy to achieve with \asO{\log n} classical communication, and therefore cannot be used as an intermediate step towards a lower bound in that model).

The main technical contribution of this paper is a way to prove an analogue of the above proposition for the more relaxed version of the communication problem.
The new proof is based on a technical lemma due to Harsha, Jain, McAllester, and Radhakrishnan~\cite{HJAR07}.

\sect{Preliminaries}
In our analysis we use the following generalization of the standard bounded error setting.
We say that a protocol solves a problem \e{with probability $\delta$ with error bounded by $\eps$} if with probability at least $\delta$ the protocol produces an answer, and whenever produced, the answer is correct with probability at least $1-\eps$.

Denote by~$\bar0$ the additive identity of a (finite) field.

Let $x=(x_1,..,x_k)$ and $y=(y_1,..,y_l)$ be tuples of sets.
We will call $x_i\cap y_j$ a \e{cell}, denoting it by $\cell(i,j)$.

Let $n$ be a power of $2$, and let us for the rest of the paper implicitly assume equivalence between any $a\in[n]$ and the lexicographically \ord[a] element of $\GF2^{\log n}$.
Define the following communication problem.
\defi{Let $x=(x_1,..,x_n)$ and $y=(y_1,..,y_n)$ be sequences of subsets of $[4n^2]$, where each subset is of size $n$.
Let $z=\big((i_1,j_1,c_1),..,(i_{t_n},j_{t_n},c_{t_n})\big)$ for $t_n\deq\floor{\log\log n}$.
Then $(x,y,z)\in \Pnn$ if it holds that \itemi{
\item For any $k_1\neq k_2$, $(i_{k_1},j_{k_2})\neq(i_{k_2},j_{k_2})$;
\item for every $k\in[t_n]$, either $\sz{\cell(i_k,j_k)}\ne2$, or $c_k\in [4n^2]\smin\set{\bar0}$ and $\lra{c_k,a+b}=0$, where $x_{i_k}\cap y_{j_k}=\set{a,b}$;
\item at least one of the following holds:
\m{\sz{\set[\sz{\cell(i_k,j_k)}=2]{k\in[t_n]}}\ge\fr{t_n}{66} \txt{~~or~~} \sz{\set[\sz{\cell(i,j)}=2]{(i,j)\in[n]\times[n]}}<\fr{n^2}{65}.}
}}

Unless stated otherwise, we will assume that input to \Pnn\ satisfies the following additional promises (which, in particular, make the problem non-trivial):\itemi{
\item at least $\fr{n^2}{65}$ cells are of size $2$;
\item $\Max[{a\in[4n^2]}]{\sz{\set[a\in x_i~\txt{or}~a\in y_i]{i}}}\le4\sqrt{\log n}$;
\item $\sum_{i,j\in{[n]}}\sz{\cell(i,j)}\le2n^2$.
}
We denote by \Unn\ the uniform distribution of such input.

Define \Unp\ to be the uniform distribution over all possible pairs of \f n-tuples of \f n-element subsets of $[4n^2]$ (observe that the distribution is \e{product} not only with respect to the two sides of input, but in fact every $x_i$ or $y_i$ is chosen independently from the rest of the input -- this property will play a crucial role in our lower bound argument).

The following estimation is straightforward.
\obs[obs_Unp]{\Unp\ produces an instance outside the support of \Unn\ with probability \aso{1}.}

In particular, \Unp\ produces trivial instances of \Pnn\ with probability that can be ignored, as long as we are concerned about solving the task with constant-bounded error.

We will give an efficient \RIIe-protocol that correctly solves \Pnn\ with probability $1-\aso1$, for \e{any} input from the support of \Unn.
Then we will show that any short deterministic \R-protocol fails with probability \asOm1, when the input is drawn from \Unp.
Thanks to \obsref{obs_Unp} and the Minimax theorem, that will imply \theoref{theo_main}.

\sect[easy_qua]{Efficient protocol for \Pnn\ in \RIIe}
Let us construct an SMP protocol that uses entanglement and delivers \asO{\log^2n\log\log n} classical bits to the referee in order to solve \Pnn\ with probability $1-\aso1$, as long as the input comes from the support of \Unn.

For any tuple of sets $x=(x_1,..,x_k)$, such that $\cup x_i\sbseq[m]$ for some $m$ being a power of $2$, we define a measurement $\Pi_x$ acting on $\log m$ qubits, as follows.

Let $\alpha_x\deq\Max[i\in{[m]}]{\sz{\set[i\in x_j]{j}}}$.
Intuitively, in defining $\Pi_x$ we would like to measure $\log m$ qubits with the $k+1$ projectors $\tilde E_i^{(x)}\deq\sum_{j\in x_i}\kbra{j}$ and $\tilde E_0^{(x)}\deq\sum_{j\nin\cup x_i}\kbra{j}$ (indeed, that will be the action of $\Pi_x$ when $\alpha_x=1$).
However, if $\alpha_x>1$ then the above collection of projectors does not constitute a valid measurement, as $\sum E_i^{(x)}\sdnle I$.
To handle that, define:
\m{\Pi_x\deq\llp E_0^{(x)}\dc E_k^{(x)}\rrp;~
E_i^{(x)}\deq\fr1{\alpha_x}\tilde E_i^{(x)} \txt{~for~} 1\le i\le k;~
E_0^{(x)}\deq I-\sum E_i^{(x)}.}
Defined this way, $\Pi_x$ is a valid quantum measurement.\fn
{Although $\Pi_x$ is not, in general, a projection, it is a measurement of a very special kind -- a \e{POVM with commuting elements}.}

Consider the following protocol $S$.
\enum{
\item Alice and Bob share, in the natural way, an entangled state $\fr1{2n}\sum_{t\in{[4n^2]}}\ket t\ket t$.
\item \label{step_2} Alice measures her part of the shared state with $\Pi_x$ and Bob measures his with $\Pi_y$.
\item \label{step_3} Alice and Bob both apply the Hadamard transform to both of the $\log n$ qubit halves of the shared state.
\item \label{step_4} Alice and Bob both measure their parts of the state in the computational basis.
\item Alice and Bob send the outcomes of both measurements to the referee (\ie $4$ outcomes are sent).
\item Upon receipt of $(i,k)$ from Alice and $(j,l)$ from Bob, the referee outputs $(i,j,k+l)$.
}
Obviously, $S$ does not solve \Pnn, though we will see how it can be used as a ``stepping stone'' for the desired protocol.

Let us analyze the action of $S$ upon an input from the support of \Unn.
Assume that $i=i_0\neq0$ and $j=j_0\neq0$.
This means that in step \ref{step_2} of $S$, Alice's measurement has resulted in $E_{i_0}^{(x)}$ and Bob's in $E_{j_0}^{(y)}$.
Then the shared state turns into
\m{\fr1{\sqrt{\sz{\cell(i_0,j_0)}}}\sum_{t\in\cell(i_0,j_0)}\ket t\ket t}
(note that unless $\cell(i_0,j_0)$ contains less than $2$ elements, its content is not locally accessible by the players, neither at this stage nor later in the protocol).

Assume that $\cell(i_0,j_0)$ contains exactly $2$ elements, $a$ and $b$.
Then after step \ref{step_3} of the protocol the shared state becomes (ignoring normalization)
\m{\sum_{k,l}\llp(-1)^{\lra{k+l,a}}+(-1)^{\lra{k+l,b}}\rrp\ket k\ket l,}
and $\ket k\ket l$ has non-zero amplitude if and only if $\lra{k+l,a}=\lra{k+l,b}$, which is equivalent to $\lra{k+l,a+b}=0$.
Moreover, each pair $k_0$, $l_0$ satisfying this condition is equally likely to become the measurement outcome in step \ref{step_4}; in particular, the value of $t_0\deq k_0+l_0$ is drawn from uniform distribution whose support is \set[\lra{t,a+b}=0]{t}.

To sum up, if $S$ produces a triple $(i_0,j_0,t_0)$, such that $i_0\neq0$, $j_0\neq0$ and $\cell(i_0,j_0)=\set{a,b}$ for some $a\neq b$, then \itemi{
\item $t_0\neq\bar0$ with probability $1-\aso1$, and
\item $\lra{t,a+b}=0$ with certainty.
}

It remains to analyze the probability that $\sz{\cell(i,j)}=2$.
The joint measurement taken by the parties in step \ref{step_2} can be written as
\m{\Pi_{x,y}\deq\set{\fr1{\alpha_x\alpha_y}\tilde E_{i,j}^{(x,y)}}_{i,j=1}^n\cup\set{E_0^{(x,y)}},}
where
\f{\tilde E_{i,j}^{(x,y)}\deq \llp\sum_{t_1\in x_i}\kbra{t_1}\rrp\otimes\llp\sum_{t_2\in y_j}\kbra{t_2}\rrp}
and $E_0^{(x,y)}\deq \llp I-\sum E_{i,j}^{(x,y)}\rrp$ (the latter corresponds to the case when $S$ returns $i=0$ or $j=0$).

Let $E_{i,j}^{(x,y)}\deq\fr1{\alpha_x\alpha_y}\tilde E_{i,j}^{(x,y)}$.
Then for any $i_0\neq0$ and $j_0\neq0$,
\m{\PR{i=i_0,j=j_0}
=\tr\llp
\fr1{2n}\sum_{s_1\in{[4n^2]}}\bra{s_1}\bra{s_1}\tm
E_{i_0,j_0}^{(x,y)}\tm
\fr1{2n}\sum_{s_2\in{[4n^2]}}\ket{s_2}\ket{s_2}\rrp\\
=\fr{\sz{\cell(i_0,j_0)}}{4n^2\alpha_x\alpha_y}.}
In particular,
\m[sz2|ijneq0]{\PRr{\sz{\cell(i,j)}=2}{i\neq0,j\neq0}
=2\tm\fr{\sz{\set[\sz{\cell(i',j')}=2]{(i',j')}}}
{\sum_{i',j'\in{[n]}}\sz{\cell(i',j')}}
\ge\fr1{65}}
and
\mal[ijneq0]{\PR{i\neq0,j\neq0}
&\ge\PR{i\neq0,j\neq0,\sz{\cell(i,j)}=2}
=2\tm\fr{\sz{\set[\sz{\cell(i',j')}=2]{(i',j')}}}{4n^2\alpha_x\alpha_y}\\
&\ge\fr1{260\alpha_x\alpha_y}
\ge\fr1{4160\log n},}
where all the inequalities follow from the fact that input belongs to the support of \Unn.

We are ready to construct a protocol for solving \Pnn.
Simply, repeat in parallel $S$ sufficiently many times in order to guarantee that, with probability $1-\aso1$, at least $t_n$ instances produce triples $(i',j',t')$ referring to pairwise different cells, consisting exclusively of non-zero elements (\ie $i',j'\neq0$ and $t'\neq\bar0$).
If that happens -- our protocol outputs $t_n$ such triples (randomly chosen from the obtained ones).
From \bref{ijneq0}, \bref{sz2|ijneq0}, and an observation that all valid non-zero triples are equally likely to show up, it is clear that \asO{\log n\log\log n} parallel copies of $S$ are sufficient.
Such protocol solves \Pnn\ with probability $1-\aso1$, and its communication complexity is \asO{\log^2n\log\log n}.

\sect[hard_cla]{Solving \Pnn\ is expensive in \R}
We will show that solving \Pnn\ in \R\ requires a protocol of cost \asOm{\fr{n^{1/4}}{\log^2n}}.
Unless stated otherwise, all the following statements are made \wrt the model \R.

In our analysis we will consider the following communication problem.
\defi{Let $x,y\sbs [4n^2]$, such that $\sz{x}=\sz{y}=n$ and $\sz{x\cap y}=2$.
Let $\sigma$ be a permutation over $[4n^2]$ and $c\in [4n^2]\smin\set{\bar0}$.
Let $x\cap y=\set{a,b}$, then \Pii\ admits two types of answers, as follows:
\itemi{
\item $(x,y,c)\in\Pii$ if $\lra{c,a+b}=0$;
\item $(x,y,(\sigma,c))\in\Pii$ if $\lra{c,\sigma(a)+\sigma(b)}=0$.
}}
Let \Uii\ be the uniform distribution of valid inputs to \Pii.

Let $S$ be a protocol for \Pnn\ of cost $k$ that can be wrong with probability at most $\eps$ (a sufficiently small constant).
Then there exists a deterministic protocol $S'$ which is wrong with probability at most $2\eps$ \wrt \Unp, as follows from \obsref{obs_Unp} and the Minimax theorem.
Unless stated otherwise, we will assume that the input is distributed according to~\Unp.

Let $\infl(x,y,i,j,c)$ be the predicate indicating that, \wrt the instance $(x,y)$ of \Pnn, it is the case that $\sz{\cell(i,j)}=2$ and $(i,j,c)$ is a part of the output by $S'$.
Define $\infl(x,y,i,j)$ to be the predicate indicating whether there exists $c$ that would satisfy $\infl(x,y,i,j,c)$.
Intuitively, $\infl(x,y,i,j)$ means that information about $\cell(i,j)$ constitutes an influential part of the output by $S'$, when the input is $(x,y)$.
For every satisfying assignment to $\infl(x,y,i,j)$ denote by $c_{i,j}^{x,y}$ the (unique) value that satisfies $\infl(x,y,i,j,c_{i,j}^{x,y})$ -- here we assume w.l.g.\ that the $t_n$ cells being addressed in the output of $S'$ are always pairwise different, as required by the definition of \Pnn.

Denote:
\mal{
p_{i,j}&\deq \PR[x,y]{\infl(x,y,i,j) \txt{~and~}(x_i,y_j,c_{i,j}^{x,y})\in\Pii},\\
q_{i,j}&\deq \PR[x,y]{\infl(x,y,i,j) \txt{~and~}(x_i,y_j,c_{i,j}^{x,y})\nin\Pii},\\
r_{i,j}&\deq p_{i,j}+q_{i,j} = \PR{\infl(x,y,i,j)}.}
Note that the correctness assumptions for $S'$ imply that (assuming $\eps\le1/3$)
\m[m_corr]{
\sum_{1\le i,j\le n}p_{i,j}\ge\fr{(1-2\eps)t_n}{66}>\fr{t_n}{200}
\txt{~~and~}
\sum_{1\le i,j\le n}q_{i,j}\le2\eps\tm t_n.}

Denote $A\deq\set[\PRr{(x_i,y_j,c_{i,j}^{x,y})\in\Pii}{\infl(x,y,i,j)}\ge1-800\eps]{(i,j)}$.
Let us see that elements of $A$ often are influential.
\clm[clm_R_1]{$\sum_{(i,j)\in A}r_{i,j}\ge\fr{t_n}{400}$.}

\prf{From \bref{m_corr},
\m{0\le\sum p_{i,j}-\fr{t_n}{400}-\fr{t_n}{400}
\le\sum p_{i,j}-\fr{t_n}{400}-\fr{\sum q_{i,j}}{800\eps},}
and $\fr{t_n}{400}\le\sum\llp p_{i,j}-\fr{q_{i,j}}{800\eps}\rrp$.
For $i,j\in[n]$, let
\mal{
p_{i,j}'&\deq\PRr{(x_i,y_j,c_{i,j}^{x,y})\in\Pii}{\infl(x,y,i,j)}
=\fr{p_{i,j}}{r_{i,j}},\\
q_{i,j}'&\deq\PRr{(x_i,y_j,c_{i,j}^{x,y})\nin\Pii}{\infl(x,y,i,j)}
=\fr{q_{i,j}}{r_{i,j}}.}
Observe that $p_{i,j}'+q_{i,j}'=1$ and $A=\set[q_{i,j}'\le800\eps]{(i,j)}$.
Therefore
\m{\fr{t_n}{400}
\le\sum_{1\le i,j\le n}r_{i,j}\llp p_{i,j}'-\fr{q_{i,j}'}{800\eps}\rrp
\le\sum_{\mac{i,j:\\q_{i,j}'\le800\eps p_{i,j}'}}r_{i,j}
\le\sum_{(i,j)\in A}r_{i,j},}
as wanted.}

We will need the following two lemmas, based on \cite{HJAR07} and \cite{G08_C_I_C_R}, correspondingly.

\newcommand{\lemhjar}{Let $\delta_1, \delta_2>0$, $k_1, k_2\in\NN$, such that a communication problem $P$ is solvable \wrt distribution $D$ by a $k_1$-round private coin protocol which produces an answer with probability at least $\delta_1$, and a produced answer is correct with probability at least $\delta_2$.
Assume also that the mutual information between the transcript of the protocol and its inputs is at most $k_2$ bits, with probability $1-\aso{\delta_1}$.

Then $P$ is solvable \wrt $D$ by a public coin protocol of communication cost \asO{k_1+k_2} which produces an answer with probability at least $\delta_1/2$ and any produced answer is correct with probability at least $\delta_2-\aso1$.}

\nlem[lem_hjmr]{(\cite{HJAR07}, \fakelemref{5.3}, reformulated)}{\lemhjar}

The statement in \cite{HJAR07} is made in different terms, for completeness we give a sketched proof of \lemref{lem_hjmr} in the Appendix.

\nlem[lem_cicrq]{(\cite{G08_C_I_C_R})}{For some absolute constant \f\delta, any public coin \R-protocol of communication cost $k$ solving \Pii\ with error bounded by \f\delta\ returns an answer with probability \asO{\fr{k^4\log^2n+k^2\log^6n}{n^2}}.}

In the definition of \pss[\Pii] analogue in \cite{G08_C_I_C_R} (denoted there by $P_{1\times1}^{\Sigma}$) different constants are used, but \lemref{lem_cicrq} follows from \cite{G08_C_I_C_R} by a straightforward reduction argument.
Besides, the definition of $P_{1\times 1}^{\Sigma}$ requires that the permutation $\sigma$ is arbitrary but fixed, instead of being a part of the output, as allowed by our \Pii; however, all the statements made in \cite{G08_C_I_C_R} regarding $P_{1\times 1}^{\Sigma}$ are also valid \wrt \Pii.~\fn
{The lower bound proof in \cite{G08_C_I_C_R} argues that a big combinatorial rectangle cannot be consistent with \e{any} answer \wrt \e{any} permutation $\sigma$, and therefore it makes no difference whether $\sigma$ is fixed ``once forever'' or ``per rectangle''.}

Let $T$ be a random variable corresponding to the transcript of $S'$ (observe that its value is uniquely determined by the input, as the protocol is deterministic).
We will consider the mutual information between $T$ and parts of the input.
For $i\in[n]$, let $s_i^{(x)}$ and $s_i^{(y)}$ be the mutual information between $T$ and, respectively, $x_i$ and $y_i$.
Let $s_{i,j}\deq s_i^{(x)}+s_j^{(y)}$.
Since \Unp\ chooses all \pl[x_i] and \pl[y_i] in a mutually independent manner, it holds that
\m{\sum_{i,j\in[n]}s_{i,j}=n\tm\llp\sum_{i\in[n]}s_i^{(x)}+\sum_{i\in[n]}s_i^{(y)}\rrp\le nk.}

\newcommand{\clmmainimplies}{From \bref{m_corr},
\m{\asOm{t_n}\ni\sum_{(i,j)\in A}p_{i,j}\le\fr1n+\sum_{\mac{(i,j)\in A\\p_{i,j}>1/n^3}}p_{i,j}\le
\fr1n+\fr{\alpha k^3\log^8n}{n^2}\tm\sum_{(i,j)\in A}s_{i,j}\le\fr{1+\alpha k^4\log^8n}n,}
where $\asOm{t_n}=\asOm{\log\log n}$.
Therefore, $k\in\asOm{\fr{n^{1/4}}{\log^2n}}$.}

\clmapp{clm_main}
{For some absolute constant $\alpha$ and any $(i,j)\in A$, it holds that $p_{i,j}\le\Max{\fr1{n^3},\fr{\alpha\tm s_{i,j}\tm k^3\log^8n}{n^2}}$.}
{Before we prove it, let us see how the claim leads to the desired lower bound.
\clmmainimplies}
{The proof can be found in the \whereproofs.
The claim leads to the desired lower bound.
\clmmainimplies}
{\appclmmain}
{W.l.g., let $(1,1)\in A$ and $p_{1,1}>\fr1{n^3}$, we want to show that for some absolute constant $\alpha$, it holds that $p_{1,1}\le\fr{\alpha\tm s_{1,1}\tm k^3\log^8n}{n^2}$.
Let $q'$ be the value defined similarly to $s_{1,1}$, but with input distribution conditioned upon $\left[\sz{\cell(1,1)}=2\right]$.
Because the condition holds with probability bounded below by a positive constant (by analogy to \obsref{obs_Unp}), $q'\in\asO{s_{1,1}}$.
Therefore, it will be sufficient to show that $p_{1,1}\in\asO{\fr{q'\tm k^3\log^8n}{n^2}}$.

We will use $S'$ as a protocol for solving \Pii\ by embedding its instance into coordinate $(1,1)$ of \Pnn, and the rest of input will play the role of private randomness.
That is, in order to solve \Pii\ with input $(x',y')$, we will run $S'$ over the pair $\big((x',r_1,..,r_{n-1})(y',r_n,..,r_{2n-2})\big)$, where $r_1,..,r_{n-1}$ are independent random subsets of $[4n^2]$ of size $n$, chosen by Alice, and $r_n,..,r_{n-2}$ are similarly chosen by Bob.
Note that our embedding requires only private randomness and does not cost any communication.

Let $l\deq\ceil{\fr k{q'}}$, and denote by $S_l'$ a private coin protocol that receives $l$ instances of \Pii\ and runs $l$ independent copies of $S'$ in parallel in a mutually independent manner, one for each instance of \Pii\ embedded into \Pnn\ as described.
After that $S_l'$ outputs $(j,c_j)$ for a random $j$, such that the \ord[j] copy of emulated $S'$ contains $(1,1,c_j)$ among its answers; if no such $j$ exists then $S_l'$ returns no answer.
The following properties of $S_l'$ with respect to $l$ input pairs independently chosen according to \Uii\ are easy to verify:\itemi{
\item If an answer is returned then $c_j$ is a right answer to the \ord[j] instance of \Pii\ with probability at least $1-800\eps$ (this follows from $(1,1)\in A$).
\item The probability that an answer is returned is $1-(1-r_{1,1})^l\ge\Min{\fr12,\fr{lp_{1,1}}2}$.
\item The protocol makes at most $k$ iterations.
\item The probability that the transcript of $S_l'$ contains more than $3q'l\sqrt{\log n}$ bits of information about the input is at most $\fr1{n^4}$ (this follows from the Chernoff bound and mutual independence of the input instances).
}

We view $S_l'$ as a protocol for solving the following distributional problem:\ \e{the input consists of $l$ independent instances of \Pii, each chosen according to \Uii, and to solve the problem one should provide a correct answer to any given instance of \Pii}.
Note that the probability that $S_l'$ returns an answer is \asOm{1/n^3} (as we assume that $p_{1,1}>\fr1{n^3}$), and therefore we can use \lemref{lem_hjmr}.
The lemma implies that there exists a public coin protocol $S''$ of cost \asO{k\log n} that receives $l$ instances of \Pii\ and returns an answer to one of them with probability at least $\Min{\fr14,\fr{kp_{1,1}}{4q'}}$, such that if an answer is returned it is correct with probability at least $1-801\eps$.

So far we have refrained from using using public randomness in our constructions, in order to be able to use \lemref{lem_hjmr} (allowing private randomness only).
Now we are going to use public randomness in order to construct a protocol $S'''$ for solving a single instance of \Pii.

If we apply a uniformly random permutation over $[4n^2]$ to any $(x,y)$ from the support of \Uii, we obtain a new pair $(x',y')$, drawn from \Uii.
Our $S'''$ will use public randomness to choose $l$ uniformly random permutations $\sigma_1\dc\sigma_l$, and apply them to its input $(x,y)$ to obtain $l$ new instances $\llp\sigma_1(x),\sigma_1(y)\rrp \dc \llp\sigma_l(x),\sigma_l(y)\rrp$.
Then $S'''$ will feed these $l$ instances to $S''$; if $S''$ outputs $(j,c_j)$ then $S'''$ will return $(\sigma_j,c_j)$, otherwise it will refuse to answer.

Like $S''$, our $S'''$ returns an answer with probability at least $\Min{\fr14,\fr{kp_{1,1}}{4q'}}$, and if an answer is returned it is correct with probability at least $1-801\eps$.
Unlike its predecessor, $S'''$ deals with a single input instance of \Pii, and therefore for sufficiently small $\eps$ \lemref{lem_cicrq} guarantees that
\m{\fr{kp_{1,1}}{4q'}\in\asO{\fr{k^4\log^6n+k^2\log^8n}{n^2}},}
which leads to $p_{1,1}\in\asO{\fr{q'k^3\log^8n}{n^2}}$, as required.}

\sect[s_open]{What we haven't done}
\itemi{
\item Can shared entanglement be stronger than \e{quantum} communication?
In fact, it is wide-open how to compare any model with entanglement to either \Q\ or \QI, due to the fact that we do not know how much entanglement can be needed to solve a communication problem of given input length.
\item Is it possible to find a \e{partial functional} problem that requires exponentially more communication in \R\ than in \RIIe?
How about a \e{total function}?
\item Is it possible to find a problem that requires exponentially more communication in \R\ than in \QII?
}

\bib

\newpage
\appendix
\sect{Appendix}

\lemrep{lem_hjmr}{\lemhjar}

\prf[\lemref{lem_hjmr}]{Let $S$ be the protocol guaranteed by the lemma requirement, assume that the input $(x,y)$ is produced according to the distribution $D$.
Let $J$ be a random variable representing the amount of mutual information between the transcript of $S$ and the input.

It is shown in \cite{HJAR07} how to build a public coin protocol $S'$ with the following properties:
\itemi{
\item For some absolute constant $\beta$, communication cost of $S'$ is upper-bounded by $\beta k_1+6J$. 
\item $S'$ refuses to answer with probability at most $1/3$ (denote this event by $E_1$). 
Otherwise it produces an answer, distributed according to the same distribution as the answer of $S$ for the given input.
}
Even if $E_1$ does not occur no answer may be returned, as $S$ is allowed to have positive probability to refuse to answer.
We will denote this second sort of refusal to answer by $E_2$ (note that this event is mutually exclusive with $E_1$).
It is clear though that an answer is produced (i.e., neither $E_1$ nor $E_2$ occurs) with probability at least $\fr{2\delta_1}3$, and if that happens then the answer is correct with probability at least $\delta_2$.

We construct a protocol $S''$, similar to $S'$ but equipped with the following ``cost control mechanism''.
If the total number of communicated bits exceeds $\beta k_1+6k_2$ then $S''$ halts and refuses to answer.
The probability that this mechanism actually stops a single run of the protocol is upper-bounded by the probability that $J>k_2$, which is the probability that $S$ exposes more than $k_2$ bits of information about its inputs.
The latter is in \aso{\delta_1} by the assumption, and therefore $S''$ returns an answer with probability $\fr{2\delta_1}3-\aso{\delta_1}>\fr{\delta_1}2$ (for sufficiently long inputs).
On the other hand, whenever an answer is returned it is correct with probability at least $\delta_2-\fr{\aso{\delta_1}}{\delta_1}=\delta_2-\aso1$, as required.}

\end{document}